\documentclass[aip,graphicx,twocolumn,10pt,apl]{revtex4}
\usepackage{graphicx} \usepackage{epstopdf}

\usepackage{afterpage}
\usepackage{times}

\begin{document}
\title{Conductivity of LaAlO$_3$/SrTiO$_3$ Interfaces made by Sputter Deposition}
\author{I. M. Dildar}
\author{D. B. Boltje}
\author{M. H. S. Hesselberth}
\affiliation{Kamerlingh Onnes Laboratorium, Leiden University, The Netherlands}
 \email[E-mail:]{aarts@physics.leidenuniv.nl}
\author{Q. Xu}
\author{H. W.  Zandbergen}
\affiliation{National Centre for High Resolution Microscopy, Kavli Institute
for Nanoscience, Delft Technical University, Lorentzweg 1, 2628 CJ Delft, the
Netherlands}
\author{S. Harkema}
\affiliation{Faculty of Science and Technology and MESA+ Institute for
Nanotechnology, University of Twente, 7500 AE Enschede, the Netherlands}
\author{J. Aarts}
\affiliation{Kamerlingh Onnes Laboratorium, Leiden University, The Netherlands}
\date{\today}
%\date{\today}
%
\begin{abstract}\noindent
We have investigated the properties of interfaces between LaAlO$_3$ films grown
on SrTiO$_3$ substrates singly terminated by TiO$_2$. We used RF sputtering in
a high-pressure oxygen atmosphere. The films are smooth, with flat surfaces.
Transmission Electron Microscopy shows atomically sharp and continuous
interfaces while EELS measurements show some slight intermixing. The elemental
ratio of La to Al measured by EDX is found to be 1.07. Importantly, we find
these interfaces to be non-conducting, indicating that the sputtered interface
is not electronically reconstructed in the way reported for films grown by
Pulsed Laser Deposition because of the different interplay between
stoichiometry, mixing and oxygen vacancies.
\end {abstract} \pacs {} \maketitle \vspace{-0.5cm}\noindent
%Introduction
%
\\\indent The formation of a two dimensional electron gas at oxide
interfaces is a field of interest since its discovery in 2004 by Ohtomo \&
Hwang \cite{hwang04}. They reported that the interface of two band insulators
LaAlO$_3$ and SrTiO$_3$ is conducting, which they ascribed to an intrinsic
doping mechanism driven by the polar-nonpolar discontinuity. That this
mechanism is important can be seen in the fact that a minimum LAO layer
thickness of 4 unit cells is needed to create the conducting interface, and
that the STO surface needs to be terminated with a TiO$_2$ layer. It was found
in subsequent work that, at least when grown by Pulsed Laser Deposition (PLD),
the interface properties strongly depend on the background oxygen pressure, and
not only intrinsic doping but also oxygen vacancies (extrinsic doping) must
play a role \cite{huijben09,herranz07,siemons07}. Moreover, cation intermixing
at  the interface was shown to play a role
\cite{kalabukhov07,willmott07,chambers10}, and in a recent study on samples
grown by Molecular Beam Epitaxy it was found that the La to Al ratio of the LAO
layer needs to be smaller than 1 in order to activate the interface conductance
\cite{warusawithana10}. This issue has not yet been addressed in PLD grown
interfaces.
\\\indent
Here we report on the growth LAO/STO interfaces by sputter deposition using
high oxygen pressure, which is a well-known deposition technique for oxide thin
films. By various characterization methods we find the LAO films smooth and the
interfaces atomically sharp, but we do not observe conductance. The La/Al ratio
is 1.07, which indicates that stoichiometry also is part of the mechanism which
yields conducting interfaces.
\begin{figure}[b]
\centering
\includegraphics[width=0.4\textwidth]{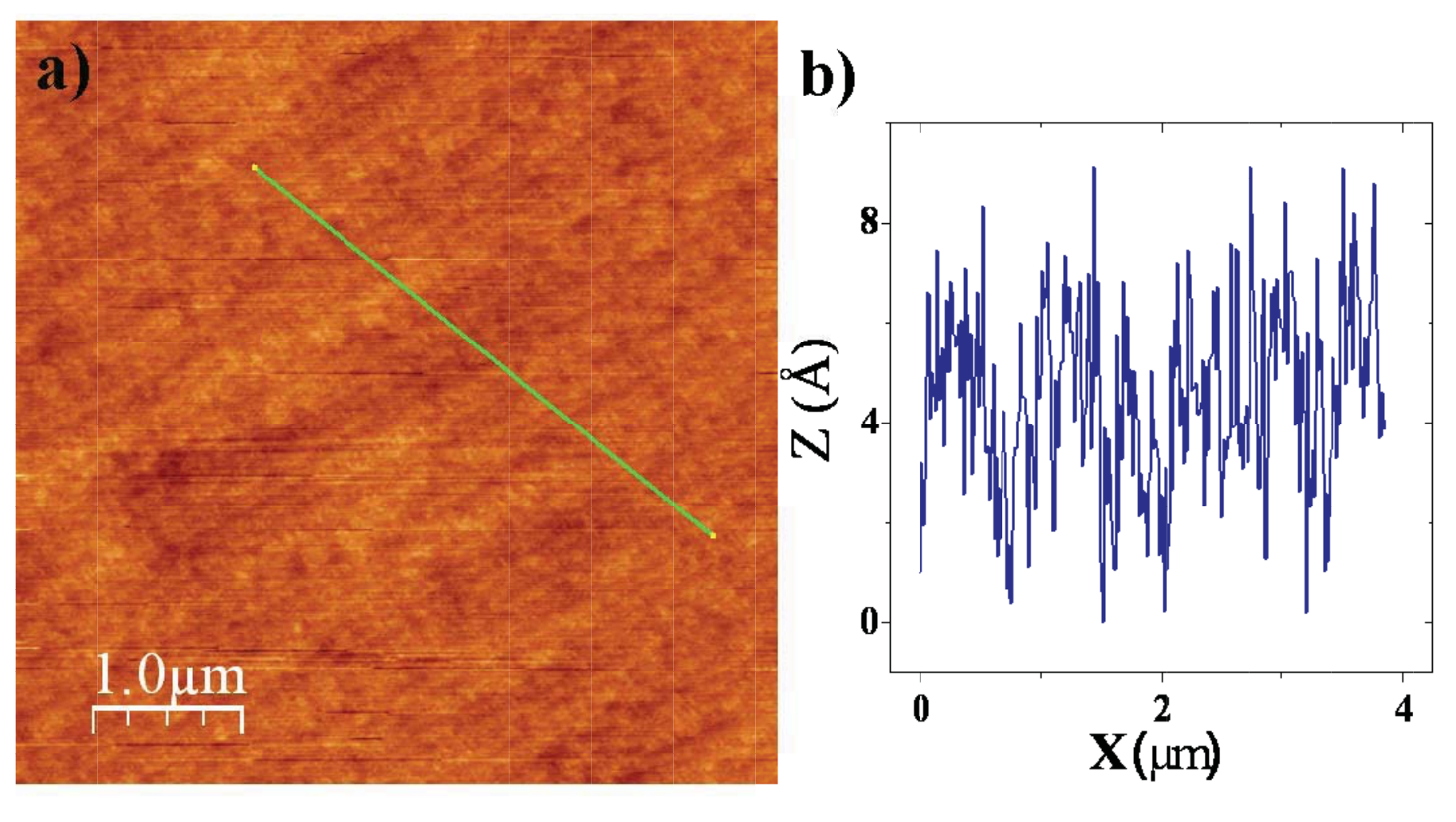}
\caption{(Color on-line) (a) Morphology of a LaALO$_3$ film on SrTiO$_3$ by
atomic force microscopy (b) Height profile along the line drawn in (a)}
\label{LAO-Fig1}
\end{figure}
%Experimentation
%
We grow thin films of LAO on the TiO$_2$-terminated surface of STO by RF
sputtering in oxygen at pressures from $0.8$ mbar to $1.2$ mbar at growth
temperatures between 900$^o$C and 940$^o$C. The morphology of the films was
characterized by Atomic Force Microscopy (AFM) in tapping mode. Thicknesses of
the grown films were measured by X-ray reflectivity (XRR) using Cu-K$\alpha$
radiation. The structural quality of the grown thin film of LAO has been
measured by X-ray diffraction (XRD). Transmission Electron Microscopy (TEM) was
used to characterize the interfaces. High Resolution TEM micrographs were
recorded using a microscope (FEI Titan cubed TEM) equipped with an image
forming Cs corrector and a High Resolution Gatan Image Filter (HR-GIF) operated
at 300~kV. Scanning TEM (STEM) was used in Energy Dispersive X-ray (EDX) mode
to determine the local stoichiometry of our LAO films. EDX profiles were
acquired on a FEI Tecnai-200 system with the probe less than 0.38 nm. The
spectra acquisition, drift correction and data analysis were all
performed using the software TIA (Tecnai Imaging \& Analysis). \\
%
%Results
%
\indent Two critical parameters which control the growth in sputtering are the
deposition temperature $T_{dp}$ and pressure $P_{dp}$. We determined a window
for smooth and epitaxial growth in Table~\ref{table:1}, which proved to be
rather narrow. Growing outside this window results in rough and structurally
defective films.

\begin{figure}[t]
\centering
\includegraphics[width=7cm]{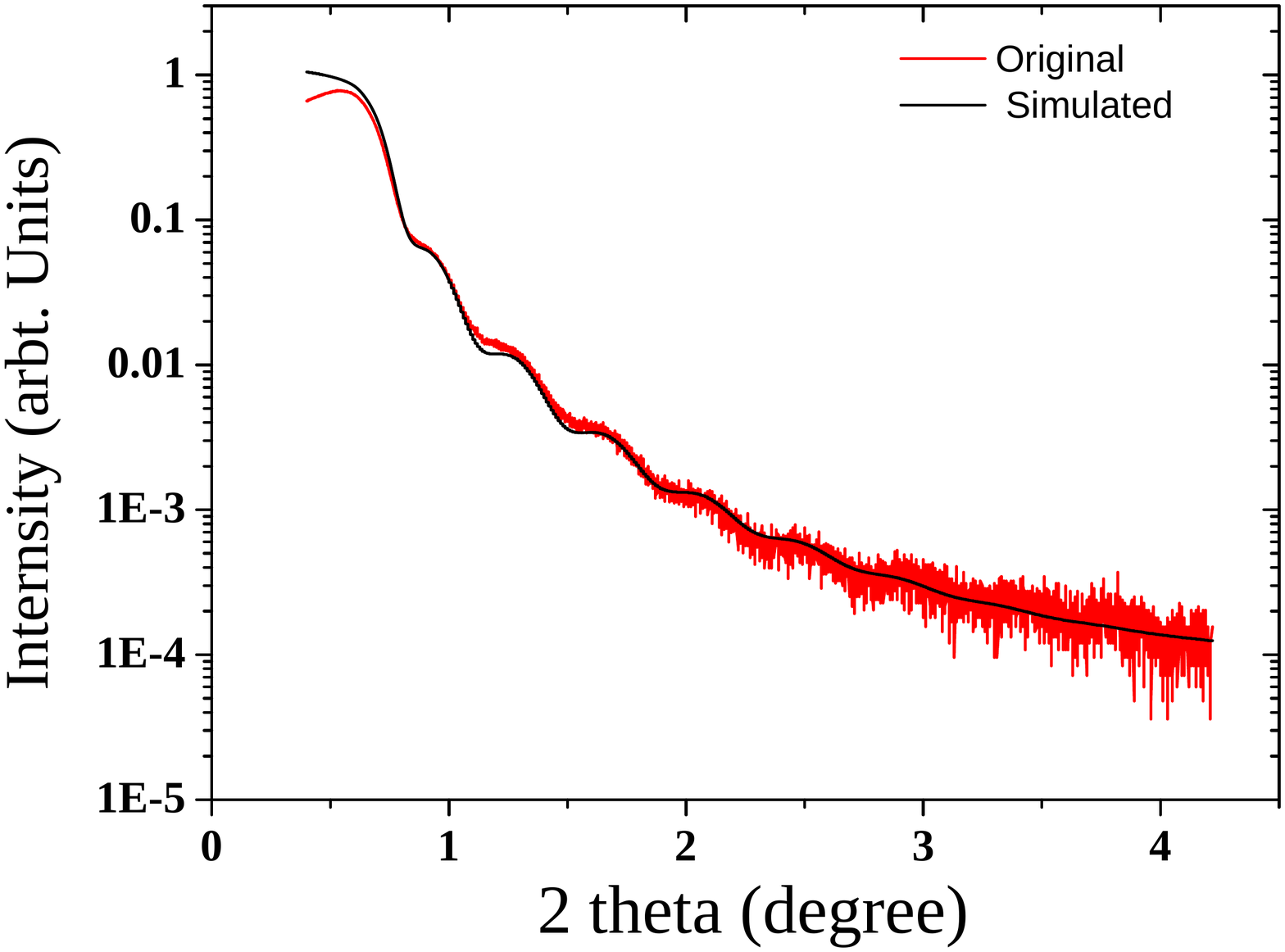}
\caption{(color on-line) X-ray reflection data for the 20~nm film LA051 (LAO on
STO). The drawn (black) line is a simulation.}\label{LAO-Fig2}
\end{figure}

\begin{figure}[t]
\includegraphics[width=7.5cm]{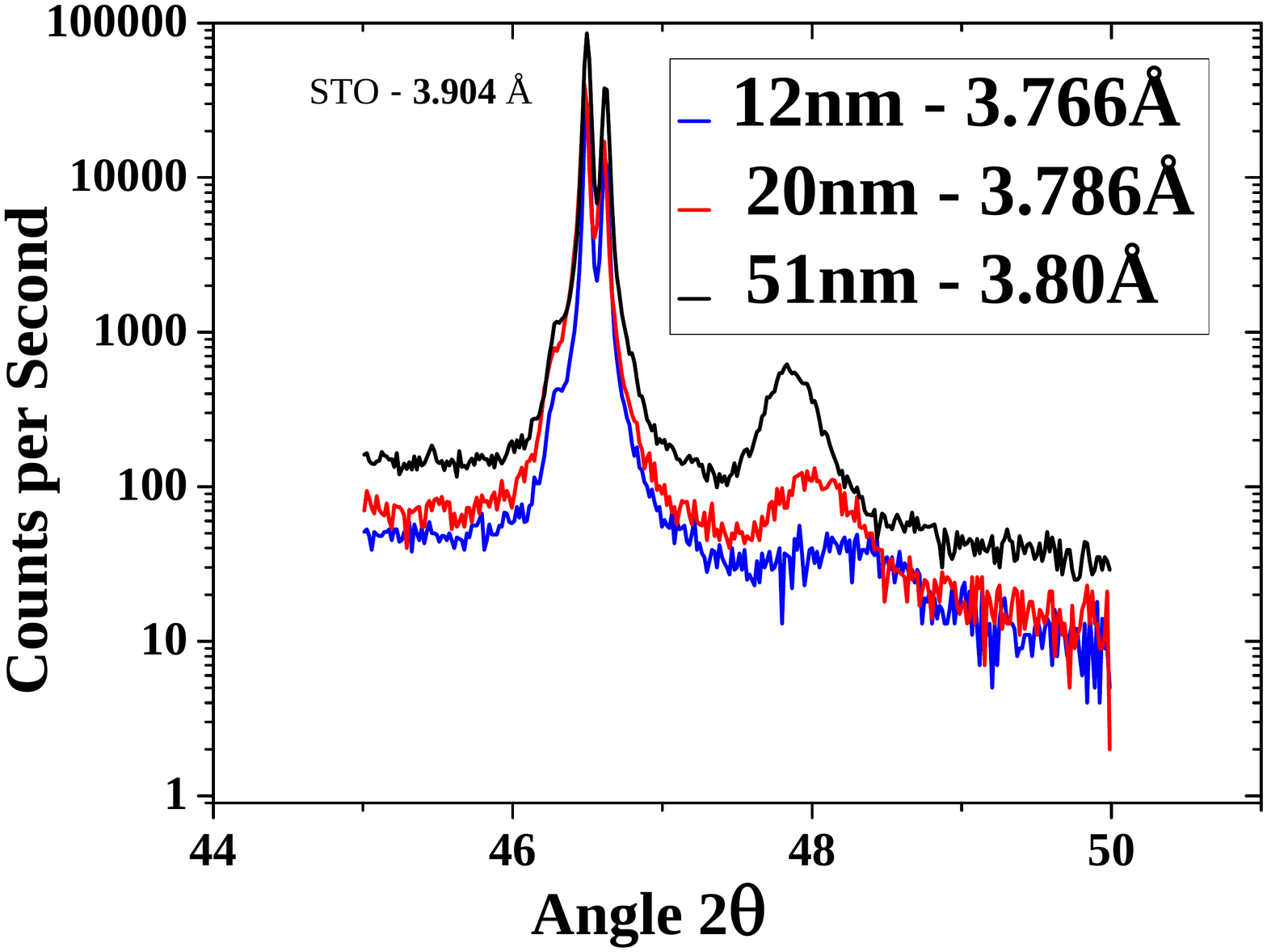}
\caption{(color on-line) XRD data of three representatives films of LAO at
angles around the (002)reflection. The (002) STO peak corresponds to a lattice
parameter of 0.3904 nm. The black line shows a 51~nm thick LAO film, the red
line is for a 20 nm film (LA051), the blue line is for a 12~nm film. Also given
are the lattice parameters of the films as calculated from the intensity
peaks.}\label{LAO-Fig3}
\end{figure}

\begin{figure}[b]
%\vspace{1.0cm}
%\includegraphics[width=3cm,height=4cm,angle=90]{Ishrat_LAO-STO_TEM-2nm.eps}
\includegraphics[width=8cm]{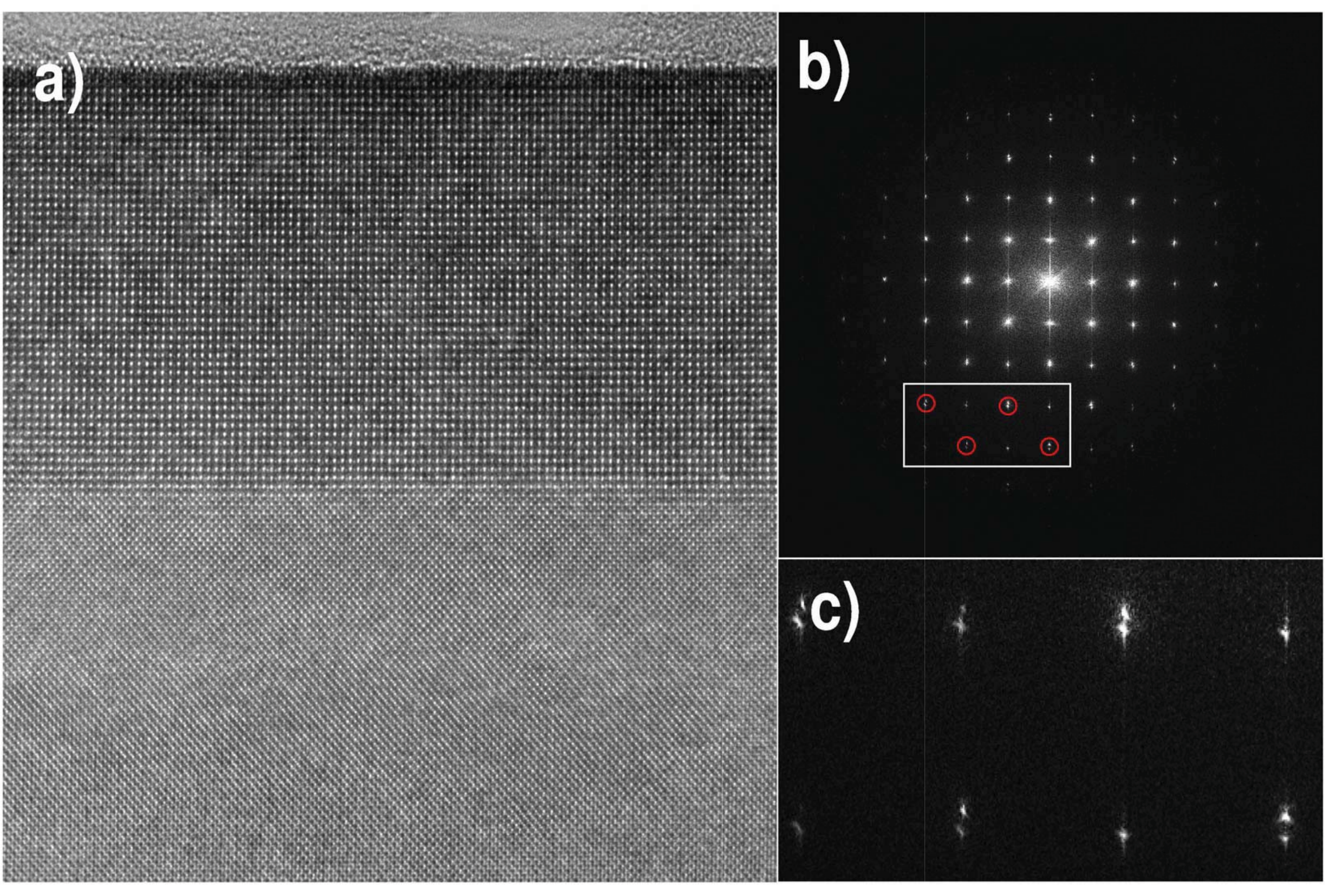}
\caption{(a) High resolution TEM picture of the LAO/STO interface for film
LA051. (b) Diffraction pattern. (c) Enlarged part of the region of the marked
spots, showing a slight splitting which indicates some misalignment between
film and substrate.}\label{LAO-Fig4}
\end{figure}

\begin{figure}[t]
\includegraphics[width=0.4\textwidth]{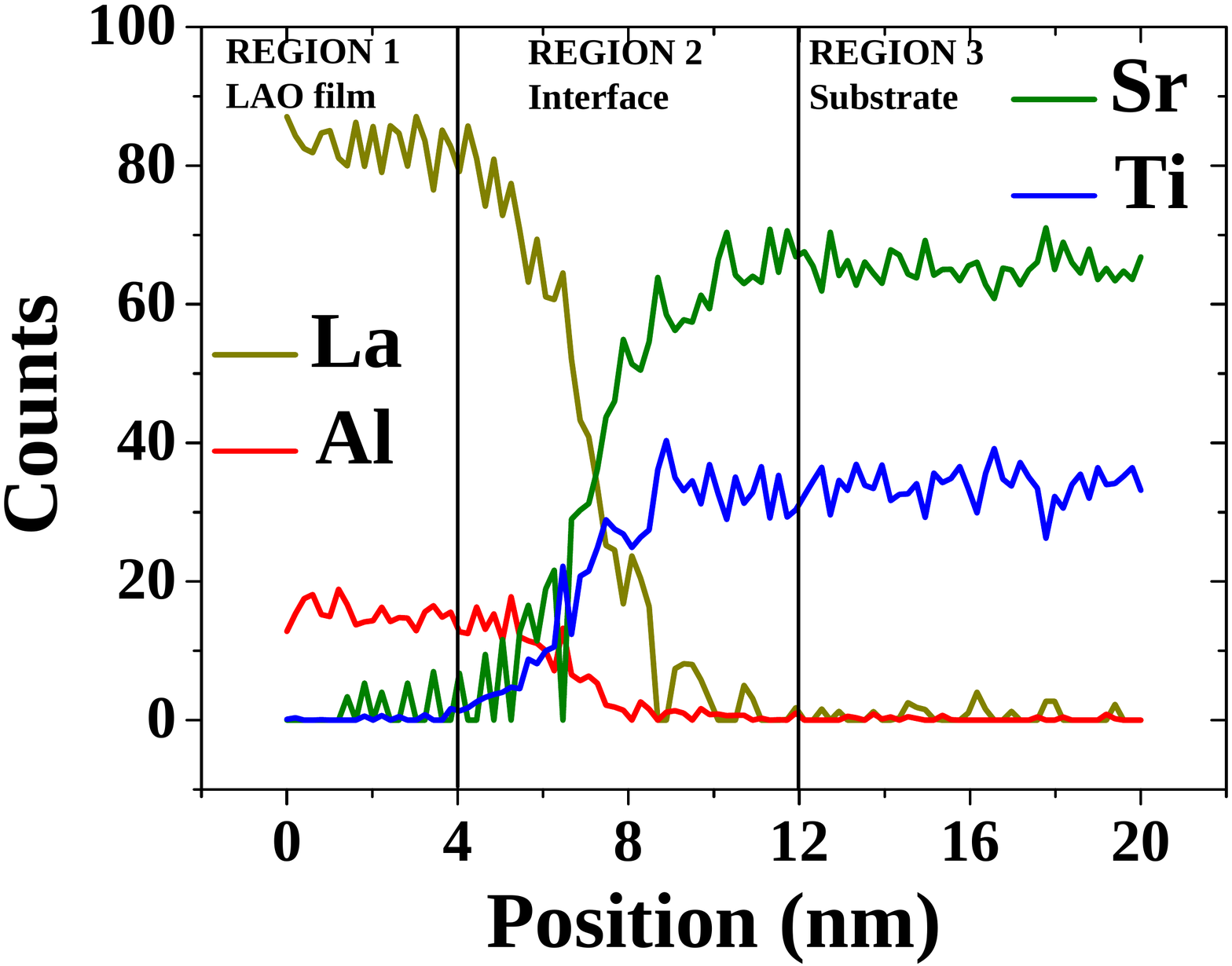}
\centering\caption{(color on-line) STEM/EDX elemental profile of the LAO/STO
sandwich across the interface. The elemental ratios La/Al and Sr/Ti were
determined in regions 1 and 3, respectively.}\label{LAO-Fig5}
\end{figure}

\begin{table}[t] \noindent
\centering
\begin{tabular}{c c c c c}
\hline\hline
$P_{dp}$(mbar) & $T_{dp}$($^oC$) & Rough.(nm) & $c_0$(\AA) & $d_{LAO}$ (nm) \\
[0.5ex] \hline\hline
1.2 & 800 & 1.6 & x  &  5 \\
1.2 & 840 & 1.7 & x  &  15 \\
1.2 & 900 & 2.1 & x   &  8  \\
1.2 & 1034 & 0.4 & 3.786  &  13  \\
1.0 & 840 & 1.4 & 3.789 &  13 \\
0.8 & 840 & 2 & 3.789  &  13  \\
0.8 & 920 & 0.2 & 3.786 & 20 \\
0.8 & 920 & 0.2 & 3.777 & 12 \\
0.6 & 940 & 0.4 & 3.799 & 14 \\
0.4 & 940 & 0.2 & x & 15 \\ [1ex]
\hline %inserts single line
\end{tabular}
\caption{Sputter deposition parameters of LaAlO$_3$ on SrTiO$_3$. Given are the
sputter gas pressure $P_{dp}$, the substrate temperature $T_{dp}$, the
roughness of the LAO film, the out-of-plane lattice parameter $c_0$, and the
LAO film thickness. The 20~nm films is LA051.} \label{table:1}
\end{table}
Good films were grown around $T_{dp}$ = 920$^oC$ and 0.8~mbar.
Figure~\ref{LAO-Fig1}a shows the surface morphology of a 20~nm film (called
LA051) measured by AFM. The corresponding profile (Fig.~\ref{LAO-Fig1}b)
indicates a step size of unit cell height i.e., 0.4~nm. The roughness of the
films is 0.2~nm over a scale of 1$\mu$m. Figure~\ref{LAO-Fig2} shows an XRR
measurement on a 20~nm film. The Kiessig fringes are clearly visible and point
to crystalline order of atomic planes perpendicular to the growth direction as
well as flatness of surface and interface. For several films, the density
profile was simulated by using Bruker XRD software. They have a constant
density for each layer which indicates homogeneous films over whole thickness
range.
\\\indent The out-of-plane lattice constant $c_0$ of the LAO films was
characterized by XRD. Figure~\ref{LAO-Fig3} shows three representative films
with thicknesses 12~nm, 20~nm (LA051) and 51~nm. The values of $c_0$ are given
in Table~\ref{table:1}. Comparison with the bulk lattice constant of LAO ($a_0$
= 3.789\AA) shows that the 12~nm film is fully strained, and the 51~nm film
fully relaxed.
Figure~\ref{LAO-Fig4}a shows a TEM micrograph of the atomically sharp LAO-STO
interface, made on film LA051. The diffractogram (Fig.~\ref{LAO-Fig4}b) shows a
small splitting in the higher order diffraction spots, which point to a small
misalignment between the out-of-plane crystallographic axes of LAO and STO.
The elemental variation across the interface was probed with EDX (beam diameter
0.2~nm) and is given in Fig.~\ref{LAO-Fig5}.
The signals are strong and stable beyond the interface region and allow to
determine the elemental composition. EDX line scans were made across the
LAO-STO interface, providing the atomic composition profiles as shown in
Fig.\ref{LAO-Fig5}. The La/Al ratio of the film was obtained by averaging over
92 data points from Region 1and calibrated by using the averaged value obtained
from a LaAlO$_3$ crystal. For an accurate calibration, the EDX experimental
conditions for the crystal and the film were deliberately set up in the same
way, including the cross-section sample preparation, the orientation of the
sample in the holder and the TEM mode settings. In this way, the La:Al ratio of
the film was found to be 7 percent higher than that of the crystal. Supposing
the ratio there to be 1, the La/Al ratio in the film is therefore 1.07. It is
of interest to note that in a recent study of growth of LAO on STO by PLD, it
was found that the out-of-plane lattice parameter $c_0$ of the LAO film is
correlated to the La/Al ratio \cite{qiao11}. A typical value of 0.378~nm for
our thin strained films (see Table~\ref{table:1}) would correspond to a La/Al
ratio of 1.10, in very good agreement with the value we find from EDX.
Finally, we determined the conductance of a number of films at room
temperature. For this, wires were bonded for a 4-point measurement, with
contacts in line. Typical values of the sheet resistance were 10~M$\Omega$ and
above. Four films were cooled down to 10~K, but showed no variation in
conductance. \\
\noindent The picture from the data is then as follows. The LAO films, and the
LAO/STO interfaces prepared by sputtering in a high oxygen pressure have
crystallographic properties very similar to those grown by PLD or MBE, but the
interface is not conducting. This points again to the important role of oxygen
vacancies, and in that respect our results bear strong resemblance to a recent
study of PLD-grown LAO/STO interfaces by Kalabukhov {\it et al.}, where the
oxygen pressure was varied between 10$^{-4}$~mbar and 5$\times 10^{-2}$~mbar
\cite{kalabukhov11}. The latter pressure is an order of magnitude higher than
where conductance, in conjunction with magnetism, is still found
\cite{huijben09}, and at this pressure the interfaces were not conducting.
Apparently, both in high-pressure PLD and in high-pressure sputtering, the
amount of oxygen vacancies produced in the growth process becomes too low to
generate a doped interface. This may not be simply due to the high gas
pressure, which might be thought to quench vacancy production by highly
energetic particles in the PLD- or sputter-plasma. The off-stoichiometry also
plays a role in the process. For instance, it was demonstrated by Schneider
{\it et al.} that oxygen is drawn out of the STO substrate in the case of LAO
films grown at low oxygen pressure (1.5~$\times 10^{-5}$ mbar), and probably
Al-rich \cite{schneider10}. Such a mechanism to create oxygen defects may not
be present in La-rich films, as was surmised by Chambers \cite{chambers11}.
Also, from first-principle density functional calculations, Hellberg concluded
that in La-rich films, La does not substitute for Al but instead, Al vacancies
are formed \cite{hellberg11}. These vacancies can migrate to the interface and
screen the polar discontinuity, so that the metallic interface does not form.
This does not answer the question whether the La-enrichment results of the high
oxygen pressure, but it does help to understand why La-rich LaAlO$_3$ on
SrTiO$_3$ does not yield
conductance. \\
\noindent In conclusion, we have grown LAO/STO interfaces by sputtering in high
oxygen pressure. The LAO films are smooth, strained for small thickness, and
show excess of La, while the interfaces are not conducting. Although sputtering
is an important deposition technique, the materials science of the LAO/STO
interfaces appears to be such that it cannot be simply utilized to produce such
two-dimensional interface conductance.
%
%Acknowledgements
%
\\\indent This research was funded through a research grant of the Stichting
FOM. I. M. Dildar is supported by the Higher Education Commission (HEC) of
Pakistan and on study leave from the Department of Physics, University of
Engineering and Technology (UET), Lahore, Pakistan.
%
%References
%

\end{document}